\documentclass[reprint,superscriptaddress]{revtex4-1}
\usepackage[utf8]{inputenc}

\usepackage{amsmath}
\usepackage{bm}
\usepackage{graphicx}
\usepackage{sidecap}
\usepackage{xcolor}
\usepackage{hyperref}

\newcommand{\p}{\partial}

\newcommand{\exx}{\bm{\hat{e}}_x}
\newcommand{\eyy}{\bm{\hat{e}}_y}
\newcommand{\ezz}{\bm{\hat{e}}_z}

\newcommand{\dx}{\mathrm{d}x}
\newcommand{\dy}{\mathrm{d}y}
\newcommand{\dV}{\mathrm{d}V}

\newcommand{\magn}{\bm{m}}

\renewcommand{\hm}{\bm{h}_{\rm m}}
\newcommand{\Heff}{\bm{F}}
\newcommand{\heff}{\bm{f}}
\newcommand{\Je}{J_{\rm e}}

\newcommand{\Exchange}{A}
\newcommand{\DM}{D}

\newcommand{\dmscaled}{\epsilon}
\newcommand{\Anisotropy}{K}
\newcommand{\anisotropy}{\kappa}
\newcommand{\Keff}{\Anisotropy_{
\rm eff}}

\newcommand{\ldw}{\ell_{\rm w}}

\newcommand{\thickness}{d_{\rm f}}

\newcommand{\Skyrmion}{Q}
\newcommand{\skyrmion}{q}

\newcommand{\Energy}{E}

\newcommand{\STnonad}{\beta}

\newcommand{\STFlow}{U}

\newcommand{\Vel}{V}
\newcommand{\Vstt}{\Vel}
\newcommand{\Vll}{\Vel_{\rm LL}}

\begin{document}

\title{Skyrmion-antiskyrmion droplets in a chiral ferromagnet}
\author{Naveen Sisodia}
\affiliation{Department of Physics, Indian Institute of Technology Delhi, Hauz Khas, New Delhi 110016, India}
\author{Pranaba Kishor Muduli}
\affiliation{Department of Physics, Indian Institute of Technology Delhi, Hauz Khas, New Delhi 110016, India}
\author{Nikos Papanicolaou}
\affiliation{Department of Physics, University of Crete, 70013 Heraklion, Crete, Greece}
\author{Stavros Komineas}
\affiliation{Department of Mathematics and Applied Mathematics, University of Crete, 70013 Heraklion, Crete, Greece}
\date{\today}

\begin{abstract}
We find numerically skyrmionic textures with skyrmion number $\Skyrmion=0$ in ferromagnets with the Dzyaloshinskii-Moriya interaction and perpendicular anisotropy.
These have the form of a skyrmion-antiskyrmion pair and may be called chiral droplets.
They are stable in an infinite film as well as in disc-shaped magnetic elements.
Droplets are found for values of the parameters close to the transition from the ferromagnetic to the spiral phase.
We study their motion under spin-transfer torque.
They move in the direction of the spin flow and, thus, their dynamics are drastically different than the Hall dynamics of the standard $\Skyrmion=1$ skyrmion.
\end{abstract}

\maketitle

\section{Introduction}
\label{eq:intro}

Since the experimental observation of skyrmions in ferromagnetic materials with the Dzyaloshinskii-Moriya (DM) interaction, a substantial amount of work has been devoted to their statics and dynamics \cite{EverschorMasellReeveKlaeui_JAP2018}.
Chiral skyrmions are topological solitons that have the same topological features as magnetic bubbles \cite{MalozemoffSlonczewski}, but the detailed features of the chiral skyrmion profile are specific to it \cite{BogdanovHubert_JMMM1994,KomineasMelcherVenakides_NL2020}.
Most work has largely focused on the axially symmetric chiral skyrmion predicted in Refs.~\cite{BogdanovYablonskii_JETP1989,BogdanovHubert_JMMM1994}.
Different crystal symmetries give rise to various types of DM interactions and these, in turn, define the kinds of skyrmions that can be stabilised \cite{BogdanovYablonskii_JETP1989,HoffmannMelcherBluegel_NatComm2017}.
In Ref.~\cite{NayakKumarParkin_Nature2017} the observation of antiskyrmions has been reported, that is, skyrmions that have a winding number opposite to that of the standard axially-symmetric skyrmions.

The dynamics of skyrmions is linked to their topology \cite{KomineasPapanicolaou_PRB2015a,PapanicolaouTomaras_NPB1991}, specifically, it depends on the topological number, usually called the skyrmion number $\Skyrmion$.
Skyrmions with $\Skyrmion \neq 0$, such as the axially symmetric skyrmions or the antiskyrmions with $\Skyrmion = \pm 1$, are called {\it topological}, while skyrmionic textures with $\Skyrmion=0$ are called {\it topologically trivial}.
Topological, $\Skyrmion \neq 0$, skyrmions are spontaneously pinned in a ferromagnetic film \cite{PapanicolaouTomaras_NPB1991,KomineasPapanicolaou_PhysD1996}.
By contrast, topologically trivial magnetic solitons propagate freely as Newtonian particles.
An example is provided by the skyrmionium \cite{BogdanovHubert_JMMM1999,LeonovMostovoy_EPJ2013,KomineasPapanicolaou_PRB2015a}, an axially-symmetric skyrmionic texture with trivial topology.

We will study $\Skyrmion=0$ solitons that can be constructed as skyrmion-antiskyrmion pairs.
In Ref.~\cite{JagannathGobelParkin_NatComm2020}, observations of topologically trivial objects in the form of skyrmion-antiskyrmion pairs in a DM material are reported.
In Ref.~\cite{RozsaPalotas_PRB2017}, $\Skyrmion=0$ textures are found numerically within a model with frustrated isotropic exchange and DM interaction and they are called ``chimera skyrmions'' due to the coexistence of skyrmion and antiskyrmion parts.
We find numerically, within a model with DM interaction, a skyrmionic texture with $\Skyrmion=0$ that has the features of a skyrmion-antiskyrmion pair.
This is an asymmetric configuration and its shape resembles that of a liquid droplet.
The skyrmion part occupies a much larger area than the antiskyrmion part.
We find that a static droplet is a stable structure in a ferromagnetic infinite film as well as in a disc-shaped element, within appropriate ranges of the model parameters.
In the case of a film, we study the dynamics of the skyrmion-antiskyrmion pair under in-plane spin-polarized current.
This is traveling in the direction of the current and presents Newtonian dynamics and no Magnus force effect.

The outline of the paper is as follows.
Sec.~\ref{sec:formulation} defines the model and the notions used in the paper.
Sec.~\ref{sec:droplet} presents the static solutions for skyrmion-antiskyrmion pairs in an infinite film.
Sec.~\ref{sec:spinTorque} studies the dynamical behavior of a skyrmion-antiskyrmion pair under spin-polarized current.
Sec.~\ref{sec:dropInDot} presents skyrmion-antiskyrmion pairs in a disc-shaped particle.
Sec.~\ref{sec:conclusions} contains our concluding remarks.

\section{Formulation}
\label{sec:formulation}

We consider a ferromagnetic film with exchange, easy-axis anisotropy perpendicular to the film, and interfacial DM interaction.
We denote the saturation magnetization by $M_s$ and the normalized magnetization vector by $\magn = (m_x,m_y,m_z)$, with $\magn^2=1$.
The magnetic energy is
\begin{equation}  \label{eq:energy0}
\begin{split}
\Energy & = \Exchange \int \left[ (\p_x \magn)^2 + (\p_y\magn)^2 \right] \dV + \Anisotropy \int (1-m_z^2)\, \dV  \\
& + \DM \int \left[ \exx\cdot (\p_y\magn\times\magn) - \eyy\cdot(\p_x\magn\times\magn) \right]\, \dV \\
 &  - \frac{1}{2}\mu_0 M_s^2 \int \hm\cdot\magn\,\dV
\end{split}
\end{equation}
where $\exx, \eyy, \ezz$ denote the unit vectors in the respective directions, $\Exchange$ is the exchange parameter, $\Anisotropy$ is the easy-axis anisotropy parameter, $\DM$ is the DM parameter, and $\hm$ is the magnetostatic field normalized to the saturation magnetization.

We will consider a thin film where the magnetostatic field is approximated as an easy-plane anisotropy term, $\hm \approx -m_z\ezz$.
Taking into account this approximation, we define the non-local part $\hm'$ of the magnetostatic field $\hm$ via the relation
\begin{equation}  \label{eq:hmprime}
    \hm = -m_z\ezz + \hm',
\end{equation}
and note that $\hm'$ tends to zero for very thin films.
Substituting Eq.~\eqref{eq:hmprime} in Eq.~\eqref{eq:energy0}, gives rise to an effective anisotropy parameter (that includes the local part of the magnetostatic field)
\begin{equation}  \label{eq:Keff}
\Keff = \Anisotropy-\frac{1}{2}\mu_0 M_s^2.
\end{equation}

The statics and dynamics of the magnetization vector are described by the Landau-Lifshitz equation derived from the energy in Eq.~\eqref{eq:energy0}.
Including a Gilbert damping term, we have
\begin{equation}  \label{eq:LLG0}
\p_t\magn = -\gamma \magn\times\Heff + \alpha\,\magn\times \p_t\magn
\end{equation}
where $\alpha$ is the damping parameter. The effective field is defined by
\begin{equation}  \label{eq:effectiveField_def}
\Heff = -\frac{1}{M_s}\frac{\delta \Energy}{\delta \magn}
\end{equation}
and it has the form
\begin{equation}  \label{eq:effectiveField0}
\begin{split}
\Heff = \mu_0 M_s & \left[ \frac{2\Exchange}{\mu_0 M_s^2} \Delta\magn + \frac{2\Keff}{\mu_0 M_s^2} m_z \ezz \right. \\
 & \left. + \frac{2\DM}{\mu_0 M_s^2} \left( \eyy\times\p_x\magn - \exx\times\p_y\magn \right) + \hm' \right].
\end{split}
\end{equation}

It is instructive to write the dimensionless form of Eq.~\eqref{eq:LLG0}.
Using $\ldw=\sqrt{\Exchange/\Keff}$ as the unit of length, we obtain the dimensionless Landau-Lifshitz-Gilbert (LLG) equation
\begin{equation} \label{eq:LLG}
\p_\tau\magn = -\magn \times \heff + \alpha \magn\times\p_\tau\magn
\end{equation}
where
\begin{equation}  \label{eq:effectiveField}
\heff = \Delta\magn + m_z \ezz + 2\epsilon \left( \eyy\times\p_x\magn - \exx\times\p_y\magn \right)
 + \frac{\hm'}{\anisotropy}
\end{equation}
and we have introduced the dimensionless DM parameter
\begin{equation}  \label{eq:parameterDMI}
\epsilon = \frac{\DM}{2\sqrt{A\Keff}}
\end{equation}
and the dimensionless anisotropy parameter
\begin{equation}  \label{eq:anisotropy}
    \anisotropy = \frac{2\Keff}{\mu_0 M_s^2}.
\end{equation}
Eq.~\eqref{eq:effectiveField} indicates that varying $\anisotropy$ amounts to tuning the effect of the magnetostatic field. 
The normalized time variable in Eq.~\eqref{eq:LLG} is
\begin{equation}  \label{eq:time}
    \tau=\frac{t}{\tau_0},\qquad
    \tau_0 = \frac{1}{\gamma\mu_0 M_s\anisotropy}.
\end{equation}

When we set $\hm'=0$ in model \eqref{eq:effectiveField}, the ground state is the spiral for $\epsilon > 2/\pi \approx 0.637$ while we have the ferromagnetic state for $\epsilon < 2/\pi$ \cite{BogdanovHubert_JMMM1994}.
Skyrmions with axial symmetry are stable excited states on the ferromagnetic state.
All magnetic configurations are characterised by the skyrmion number defined as
\begin{equation}  \label{eq:skyrmionNumber}
\Skyrmion = \frac{1}{4\pi} \int \skyrmion\, \dx\dy,\qquad 
\skyrmion = \magn\cdot(\p_y\magn\times \p_x\magn)
\end{equation}
where $\skyrmion$ is a {\it topological density}.
The skyrmion number $\Skyrmion$ is integer-valued for all magnetic configurations with a constant $\magn$ at spatial infinity.
For definiteness, we assume $\magn=(0,0,1)$ at spatial infinity in all our calculations.

Skyrmion configurations often have a simple representation when we use the stereographic projection of the magnetization, defined by
\begin{equation}
\Omega = \frac{m_x+i m_y}{1+m_z}.
\end{equation}
For the model with the exchange interaction only (pure exchange model), an axially symmetric skyrmion solution of unit degree, $\Skyrmion=1$, is given by \cite{BelavinPolyakov_JETP1975}
\begin{equation}  \label{eq:BPskyrmion}
\Omega = \frac{a}{\rho}\, e^{i\phi} = \frac{a}{x-iy},
\end{equation}
where $(\rho, \phi)$ are polar coordinates and $a$ is an arbitrary constant giving the skyrmion radius.
The solution of the pure exchange model
\begin{equation}  \label{eq:BPantiskyrmion}
\Omega = \frac{a}{\rho}\, e^{-i\phi} = \frac{a}{x+iy}
\end{equation}
presents opposite winding than solution \eqref{eq:BPskyrmion} as seen in the sign of the complex exponent.
Such a configuration has skyrmion number $\Skyrmion=-1$ and it is called an {\it antiskyrmion}.

\section{Skyrmion-antiskyrmion textures}
\label{sec:droplet}

\label{sec:dropInFilm}

\begin{figure}[t]
\begin{center}
    \includegraphics[width=\columnwidth]{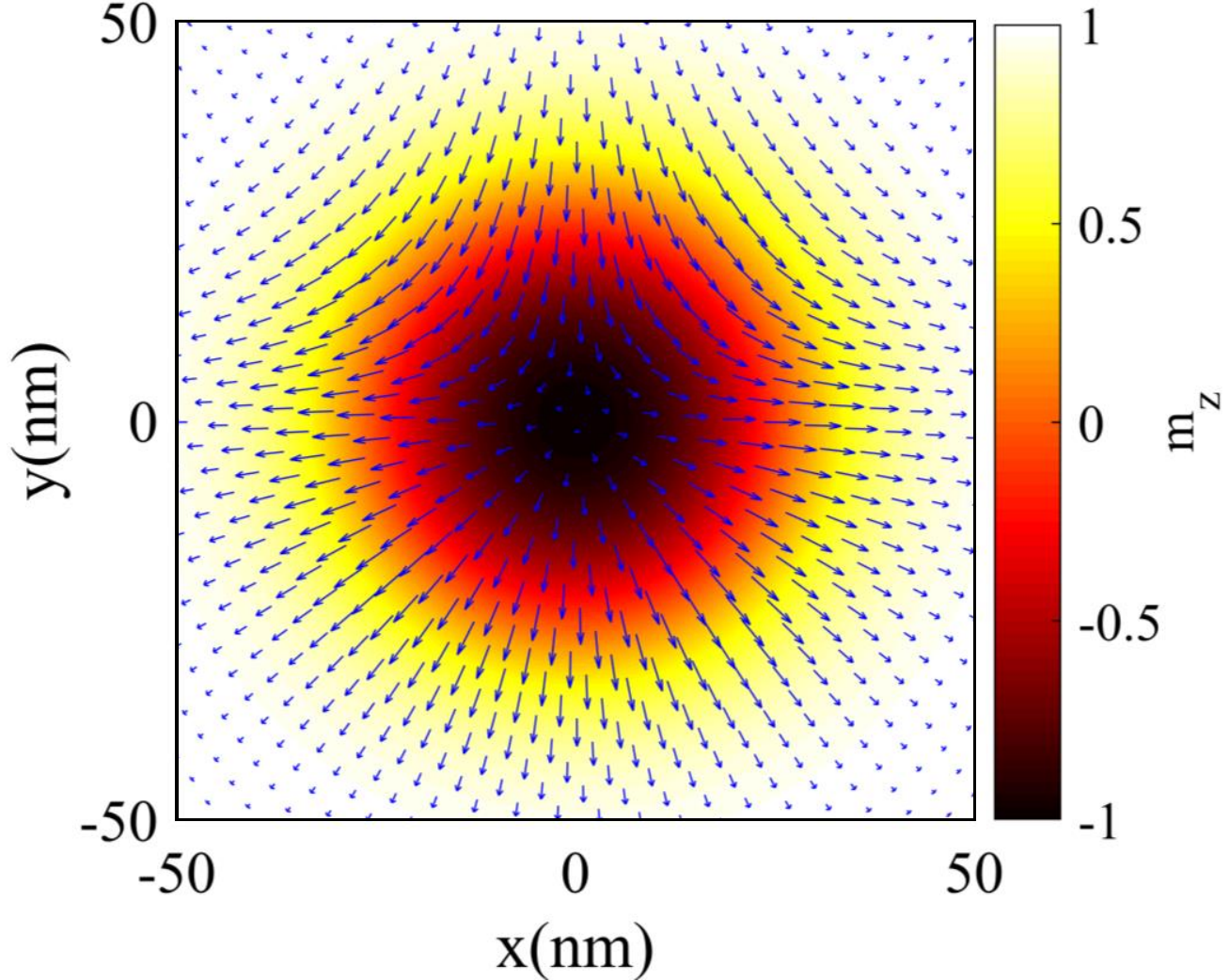}
    \caption{The magnetization configuration produced by the form \eqref{eq:BP0} representing a skyrmion-antiskyrmion pair.
    The lower half of this configuration has the features of a skyrmion and the upper half has the features of an antiskyrmion.
    This is used as an initial condition in the energy relaxation algorithm for finding a static solution of the equation.}
    \label{fig:BP0}
\end{center}
\end{figure}

We are looking for solutions of model \eqref{eq:LLG} with skyrmion number $\Skyrmion=0$.
An ansatz for a $\Skyrmion=0$ configuration is conveniently given in terms of the stereographic variable as
\begin{equation}  \label{eq:BP0}
\Omega = \frac{a}{x + i |y|}
\end{equation}
where $a$ is an arbitrary constant.
The magnetization configuration produced by the form \eqref{eq:BP0} is shown in Fig.~\ref{fig:BP0}.
Half of this configuration has the features of a skyrmion similar to the form \eqref{eq:BPskyrmion}, and the other half has the features of an antiskyrmion similar to the form \eqref{eq:BPantiskyrmion}.
Such a configuration may be called a {\it skyrmion-antiskyrmion} pair.

We perform numerical simulations using \textit{Mumax3} \cite{VansteenkisteWaeyenberge_AIPadv2014}.
We use the form \eqref{eq:BP0} as an initial condition and apply an energy minimization procedure.
First, we use the ``minimize()" function of \textit{Mumax3} that applies a conjugate gradient method for energy minimization.
The energy is minimized until the error in the magnetization is smaller than $10^{-5}$ in every micromagnetic cell.
We then integrate the Landau-Lifshitz-Gilbert equation without the precession term till either the total energy of the system reaches the numerical noise floor of the simulation or the total simulation time exceeds 10~ns.
The above methodology converges to a static skyrmion-antiskyrmion configuration for a narrow range of values of the dimensionless parameters $\dmscaled, \anisotropy$.
The convergence of the algorithm also depends on the film thickness.

\begin{table}[t]
\centering 
\begin{tabular}{c c}
\hline
Parameter & Value \\
\hline
$M_s$ & $8.38\times 10^5\,{\rm A/m}$ \\
A & $1.1\times 10^{-11}\,{\rm J/m}$ \\
$\Anisotropy (\Keff)$ & $1.193\, (7.518) \times 10^6\,{\rm J/m^3}$ \\
\DM & $3.5\times 10^{-3}\,{\rm J/m^2}$ \\
\hline
\end{tabular}
\caption{Values for material parameters used in most of our simulations presented in the figures: $M_s$ is the saturation magnetization, $A$ the exchange parameter, $\Anisotropy$ the easy-axis anisotropy parameter ($\Keff$ includes the local effect of the magnetostatic interaction according to Eq.~\eqref{eq:Keff}), and $\DM$ the DM parameter.}
\label{tab:parameters}
\end{table}

In most of the numerical simulations presented in the following figures, we use the set of parameter values shown in Table~\ref{tab:parameters}.
The resulting values for the length and time scales in Eq.~\eqref{eq:LLG} are $\ldw=3.81\,{\rm nm}$ and $\tau_0 = 5.14\,{\rm ps}$ respectively.
The values of the dimensionless parameters are
\begin{equation}  \label{eq:parameters_values}
\dmscaled=0.6086,\qquad \anisotropy=1.704.
\end{equation}

\begin{figure}[t]
    \centering
    \includegraphics[width=\columnwidth]{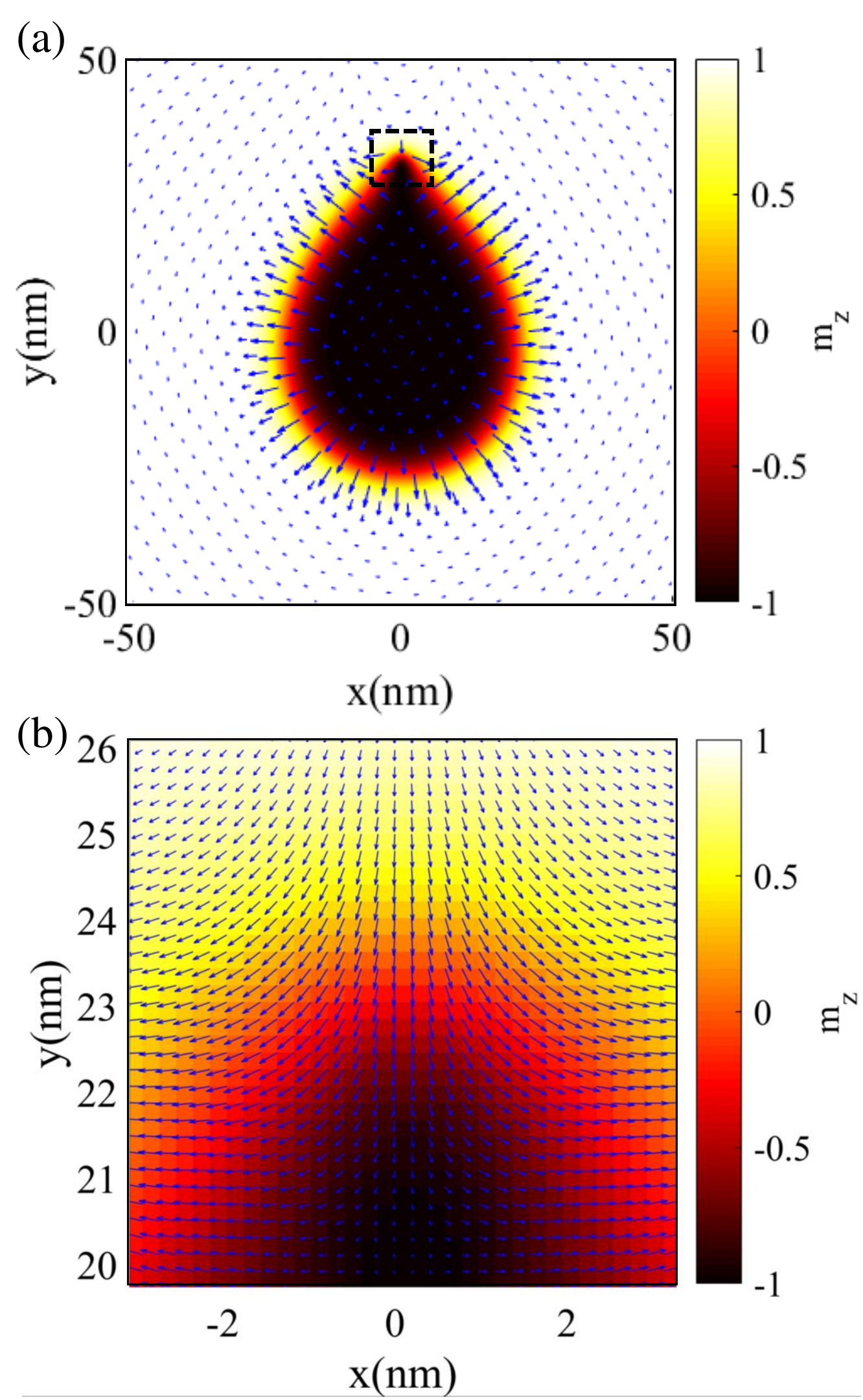}
    \caption{(a) A static skyrmion-antiskyrmion droplet with $\Skyrmion=0$ found numerically for the parameter values given in Table~\ref{tab:parameters} and for film thickness $\thickness=0.5$~nm.
    For comparison, a $\Skyrmion=1$ skyrmion in the same system has radius $\sim 16\,{\rm nm}$
    (b) Blow-up of the antiskyrmion part of the configuration, where we show the full resolution of the simulation.
    The cell size in the simulations is $\rm 0.195~nm\times0.195~nm\times0.5~nm$.
    The simulated domain in the plane of the film is $400\,{\rm nm}\times400\,{\rm nm}$ and periodic boundary conditions are used.
    \label{fig:staticDrop}}
\end{figure}

Fig.~\ref{fig:staticDrop} shows a skyrmion-antiskyrmion configuration which is found as a static solution of the LLG equation \eqref{eq:LLG0}.
This resembles, in its overall shape, a droplet of liquid and we will therefore refer to it as {\it a chiral droplet} (or, simply, a droplet).
A larger part of the configuration has the features of a skyrmion and a smaller part has the features of an antiskyrmion.
It neither is axially-symmetric nor does it have a circular shape.
In the example shown in Fig.~\ref{fig:staticDrop}, the antiskyrmion part is at the top part of the droplet, but the configuration can be rotated without changing its energy.

Textures with skyrmion and antiskyrmion parts (``chimera skyrmions''),  similar to the ones reported here, have been found numerically and studied in Refs.~\cite{RozsaPalotas_PRB2017,PalotasRozsa_PRB2017,RozsaNowak_arXiv2020} within a model with frustrated isotropic exchange interaction and DM interaction.
Related to our droplet are also the $\Skyrmion=0$ magnetic bubbles reported in Ref.~\cite{MoutafisKomineas_PRB2009}.
The latter are not chiral, they are stabilized primarily by the magnetostatic interaction and their overall shape is almost circular.
Another related structure (also termed a droplet) has been reported in Ref.~\cite{MohseniSaniAkerman_Science2013}, in films without DM interaction, and it is a dynamical configuration exhibiting precession of spins.
In the chiral droplet studied here, spin precession does not occur due to the breaking of rotational symmetry in the magnetization space introduced by the chiral interaction.
In Ref.~\cite{Cooper_PRL1998}, Skyrmion-antiskyrmion pairs have been studied for a model with exchange interaction only and these were found to be necessarily propagating.

\begin{figure}[t]
    \centering
    \includegraphics[width=0.9\columnwidth]{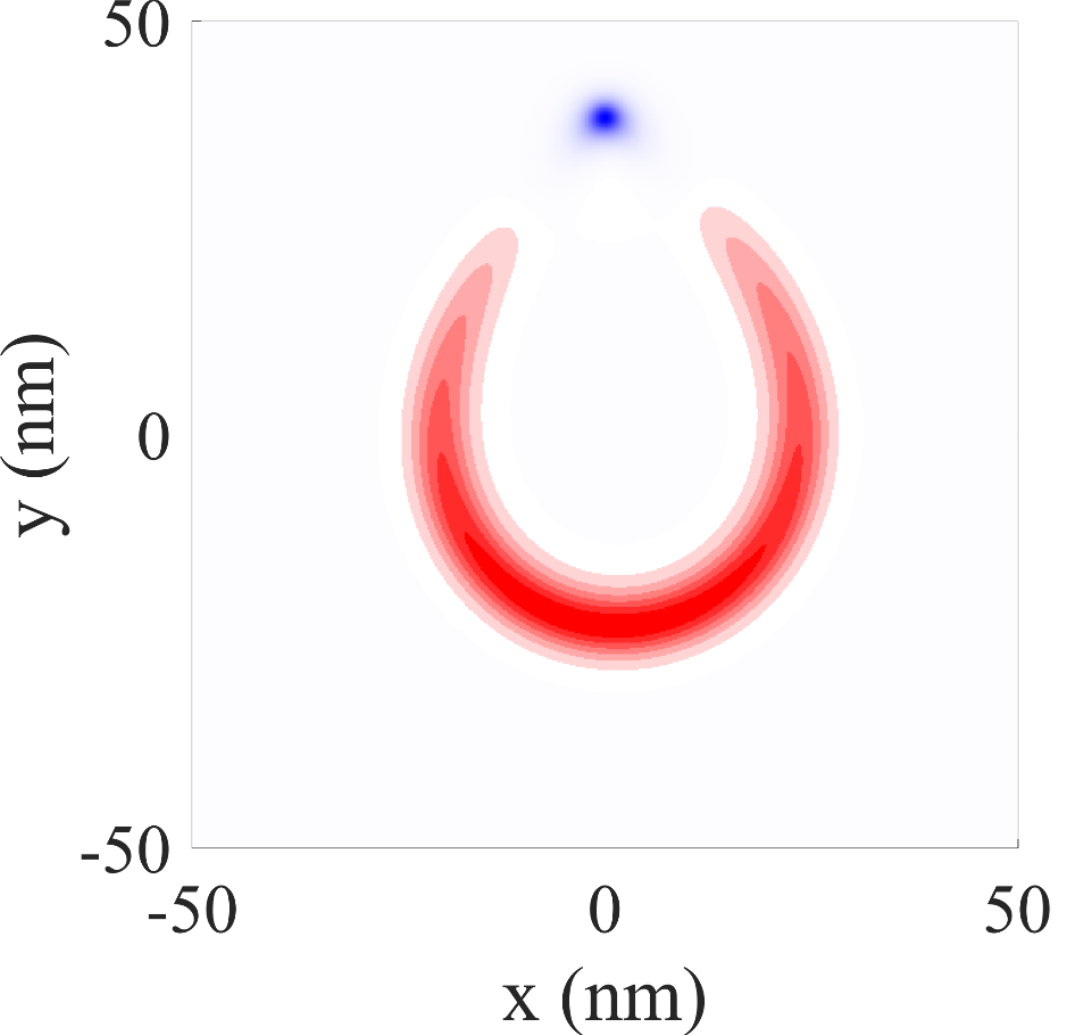}
    \caption{The topological density $\skyrmion$ for the chiral droplet of Fig.~\ref{fig:staticDrop}.
    The part where $\skyrmion<0$ (blue) occupies a very small region compared to the part where $\skyrmion>0$ (red).
    The density $\skyrmion$ takes very large negative values in a small region while, in the region with positive values, $\skyrmion$ is small.
    The integrated topological density, or $\Skyrmion$, is zero.
    \label{fig:topologicalDensity}}
\end{figure}

Fig.~\ref{fig:topologicalDensity} shows the distribution of the topological density $\skyrmion$ defined in Eq.~\eqref{eq:skyrmionNumber} for the skyrmion of Fig.~\ref{fig:staticDrop}.
The area of negative topological density is concentrated in the small part of the droplet where the antiskyrmion is located and it takes very high values.
The area of positive $\skyrmion$ is spread over a much larger area around the droplet domain wall and it takes smaller values.

We could find stable skyrmion-antiskyrmion droplets only in thin films.
For $\thickness=0.5\,{\rm nm}$, we find a static droplet for values of the dimensionless parameter in a narrow range around $\dmscaled=0.61$.
Keeping $\dmscaled$ constant, we could choose the DM parameter in the range $2.9\,{\rm mJ/m^2} \leq \DM \leq 6\,{\rm mJ/m^2}$. 
We could also find stable droplets for film thickness smaller than $\thickness=0.5\,{\rm nm}$ and for similar parameter values.
We could also find a stable droplet for film thickness $\thickness=1\,{\rm nm}$ for the parameter values $\DM=6\,{\rm mJ/m^2}, \Anisotropy=26.5\times10^5\,{\rm J/m^2}$ (other parameters as in Table~\ref{tab:parameters}).
These values correspond to the dimensionless parameters $\dmscaled=0.6086,\,\anisotropy=5.01$.
The droplet is stable for a range of parameter values around the ones given above.
We finally note that, as the parameter space is large and hard to explore exhaustively, we cannot exclude that the droplet may also exist for values of the parameters beyond those reported in this paper.

The presence of the magnetostatic field is important for the existence of a skyrmion-antiskyrmion droplet.
No such solution is found if $\hm$ is not included in the LLG Eq.~\eqref{eq:LLG}.

\section{Motion under spin-transfer torque}
\label{sec:spinTorque}

We probe the dynamics of the $\Skyrmion=0$ droplet by applying an in-plane current flowing in the magnetic film.
We model this system via the LLG equation including spin-transfer torque terms \cite{ZhangLi_PRL2004}
\begin{equation}  \label{eq:llg_stt_ip}
(\p_t + \STFlow_\mu\, \p_\mu)\magn = -\gamma \magn\times\Heff + \magn\times \left( \alpha\p_t + \STnonad \STFlow_\mu\, \p_\mu \right) \magn
\end{equation}
where we have used the notation $x_\mu$ with $\mu=1$ or $2$ for the two coordinates in the film plane.
The velocity of the spin-polarized electron flow is $(\STFlow_1, \STFlow_2)$ and we will consider the two cases $(\STFlow_1, \STFlow_2)=(\STFlow,0)$ and $(\STFlow_1, \STFlow_2)=(0,\STFlow)$, i.e., a current flowing in the $x$ and in the $y$ direction, respectively.
The flow velocity $\STFlow$ is called the adiabatic spin torque parameter and it is given by
\begin{equation}
\STFlow = \frac{P g\mu_B}{2|e| M_s}\,\Je
\end{equation}
where $\Je$ is the current density, $P$ is the degree of polarization, $\mu_B$ is the Bohr magneton, and $g=2$ is the gyromagnetic ratio. 
The parameter $\STnonad$, called the degree of adiabaticity, represents the contribution of the non-adiabatic spin torque term relative to the adiabatic one.

If we assume rigid translational motion of the droplet with a velocity $\bm{\Vstt}=(\Vel_1,\Vel_2)$, i.e., we make the traveling wave ansatz, then we have $\p_t\magn = -\Vel_\nu \p_\nu\magn$.
We substitute this in Eq.~\eqref{eq:llg_stt_ip}, take the cross product of both sides with $\p_\lambda\magn$, then contract with $\magn$, and integrate the resulting equations for $\lambda=1, 2$ over all space and set $\Skyrmion=0$ \cite{KomineasPapanicolaou_PRB2015b}.
This obtains that the motion is in the direction of the current flow with velocity \cite{EverschorGarst_PRB2011,KomineasPapanicolaou_PRB2015b}
\begin{equation}  \label{eq:vel_rigidMotion}
    \Vstt = \frac{\STnonad}{\alpha} \STFlow.
\end{equation}
Therefore, in a steady state motion, the droplet is not expected to exhibit a component of the motion perpendicular to the current, in contrast to the typical dynamics of the $\Skyrmion=1$ skyrmion.

\begin{figure*}[t]
    \centering
    \includegraphics[width=2\columnwidth]{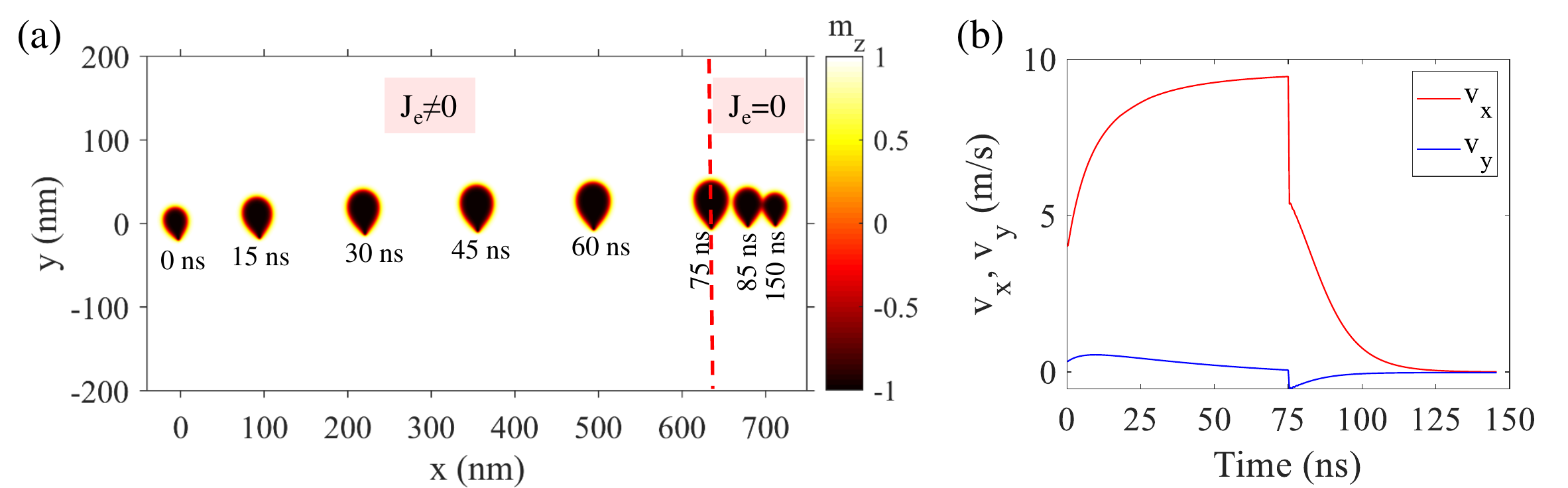}
    \caption{(a) Snapshots of an initially static droplet during a simulation where a spin current $\Je$ is applied in the $x$ direction for times $0\leq t\leq 75\,{\rm ns}$.
    The spin current parameters are $\STFlow=3.86\,{\rm m/s},\, \STnonad=0.075$ and the damping parameter is $\alpha=0.03$.
    (b) The velocity of the droplet as a function of time.
    Upon switching on the current the velocity is $\Vstt=4.02\,{\rm m/s}$, in the current direction (denoted $\Vel_x$ in the figure).
    The droplet is then accelerated up to $\Vstt=9.45\,{\rm m/s}$, which is close to the value $2.5\STFlow$.
    After switching off the current, the velocity drops instantly by approximately $4\,{\rm m/s}$ (which is close to $\STFlow$), to $\Vstt=5.37\,{\rm m/s}$.
    The droplet continues to travel and the damping term decelerates the motion until it stops.
    The component of the velocity perpendicular to the current (denoted $\Vel_y$ in the figure) is very small and goes to zero for steady-state motion.
    }
    \label{fig:STsnapshots}
\end{figure*}

For a more detailed description of the droplet motion and of the following simulations, we recall a fundamental result given in Ref.~\cite{KomineasPapanicolaou_PRB2015b}.
That is, a propagating solution of Eq.~\eqref{eq:llg_stt_ip} with velocity $\Vstt$ is also a solitary wave solution of the conservative Landau-Lifshitz equation, i.e., Eq.~\eqref{eq:LLG0} with $\alpha=0$, albeit with a different velocity $\Vll$.
Specifically, let us assume an electron flow velocity $(\STFlow,0)$ and a droplet propagating with velocity $\Vstt$ in the direction of the current, i.e., $\magn=\magn(x-\Vstt t,y)$.
Eq.~\eqref{eq:llg_stt_ip} gives
\begin{equation}  \label{eq:llg_stt_ip_x}
(\STFlow - \Vstt)\, \p_x\magn = -\gamma\magn\times\Heff + \alpha \left( \frac{\STnonad}{\alpha} \STFlow - \Vstt \right)\magn\times \p_x\magn.
\end{equation}
If we assume a propagating solution of Eq.~\eqref{eq:llg_stt_ip_x} with velocity $\Vstt = \frac{\STnonad}{\alpha}\STFlow$, then the same configuration is a solitary wave satisfying the conservative ($\alpha=0$) Landau-Lifshitz equation \eqref{eq:LLG0} with a velocity
\begin{equation} \label{eq:velocity_reduction}
\Vll = \Vstt - \STFlow = \left( \frac{\STnonad}{\alpha}-1 \right)\STFlow.
\end{equation}
In the special case $\STnonad=\alpha$, a static solution of Eq.~\eqref{eq:LLG0}, say $\magn_0(x,y)$, gives the propagating solution $\magn(x,y,t)=\magn_0(x-\STFlow t,y)$ of Eq.~\eqref{eq:llg_stt_ip_x}, with velocity $\Vstt=\STFlow$.

We now proceed to numerical simulations where we use as initial condition the static droplet shown in Fig.~\ref{fig:staticDrop}.
For the results presented in this section, we use a domain $400\,{\rm nm}\times 400\,{\rm nm}$ with periodic boundary conditions.
The cells have dimensions $\rm 0.39~nm\times0.39~nm\times0.5~nm$, i.e., it is a coarser lattice than the one used for the achievement of the static droplet in Fig.~\ref{fig:staticDrop}.
We use $P=0.56$, $g=2$ and, for our typical choice of current $\Je=1.0\times 10^{11}\,{\rm A/m^2}$, we have a flow velocity
$\STFlow = 3.86\,{\rm m/sec}$.

For following the dynamics of the droplet, we measure its position $(X,Y)$ using the formulae
\begin{equation}
    X = \frac{\int x (m_z-1)\,dx dy}{\int (m_z-1)\,dx dy},\quad
    Y = \frac{\int y (m_z-1)\,dx dy}{\int (m_z-1)\,dx dy}.
\end{equation}
We calculate the skyrmion velocity using finite differences of the position.

In the following simulations, we have chosen a current in the $x$ direction and a damping parameter $\alpha=0.03$.
We use as initial condition the droplet $\magn_0(x,y)$ of Fig.~\ref{fig:staticDrop} rotated by $\pi$ (the reason for the rotation will become apparent in the following).
For $\STnonad=\alpha$ we observe that the droplet of the initial condition is traveling with velocity $\Vstt=\STFlow$ (within numerical error) in the direction of the current, $\magn=\magn_0(x-\STFlow t,y)$, as anticipated from the discussion in connection with Eq.~\eqref{eq:velocity_reduction}.
During the simulation the initial droplet remains unchanged.
We also observe a small component of the velocity $\sim0.15\,{\rm m/sec}$ perpendicular to the current direction, and we attribute it to numerical errors.

\begin{figure*}[t]
    \centering
    \includegraphics[width=1.95\columnwidth]{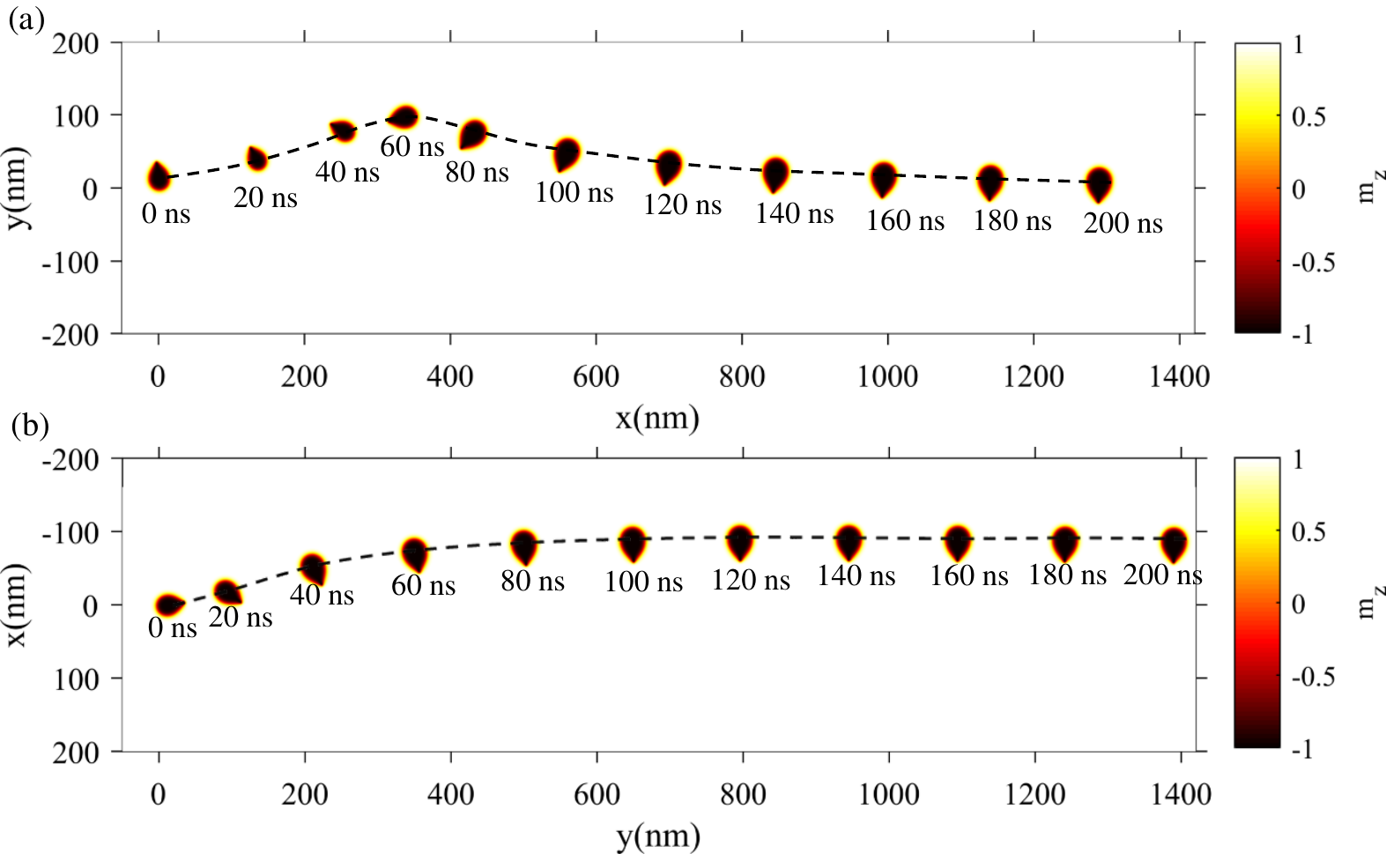}
    
    \caption{Snapshots for the droplet of Fig.~\ref{fig:staticDrop} when this is placed under spin torque at time $t=0$ and it is set in motion.
    The flow velocity is (a) $(\STFlow,0)$ and (b) $(0,\STFlow)$ with $\STFlow=3.72\,{\rm m/sec}$.
    The damping parameter is $\alpha=0.1$ and the non-adiabaticity parameter is $\STnonad=0.2$.
    The simulation domain is $400\,{\rm nm}\times 400\,{\rm nm}$ and we apply periodic boundary conditions.
    }
    \label{fig:STsnapshots_pi}
\end{figure*}

In the next simulation, we choose $\STnonad=0.075$ (that is, $\STnonad=2.5\alpha$).
A spin current is applied for the time interval $0 \leq t \leq 75\,{\rm ns}$ and it is then switched off.
Fig.~\ref{fig:STsnapshots} shows snapshots of the droplet during the simulation and the velocity of the droplet as a function of time.
Upon switching on the current the droplet instantly acquires a velocity $\Vstt$
in the current direction, which is close to $\STFlow$.
It is subsequently accelerated up to $\Vstt=9.45\,{\rm m/s}$, which is close to the value $2.5\STFlow$.
At this point, the velocity seems to saturate.
The propagating droplet is different (larger) than the static one as clearly seen in the snapshots.

When the current is switched off, at $t=75\,{\rm ns}$ the droplet velocity is reduced instantly by approximately $4\,{\rm m/s}$ (which is close to $\STFlow$), to $\Vstt=5.37\,{\rm m/s}$.
From this point on, the relevant equation is Eq.~\eqref{eq:LLG0} while the reduction of the velocity is anticipated based on Eq.~\eqref{eq:velocity_reduction}.
Let us summarize the procedure.
The spin current initially accelerates the droplet and the configuration converges to a solitary wave solution of the conservative Landau-Lifshitz equation \cite{DoeringMelcher_CVPDE2017}.
The solitary wave continues to travel in the absence of the current.
The damping term decelerates the motion until it eventually stops.
We measure a small component of the velocity in the direction perpendicular to the current ($\Vel_y$) during the acceleration phase.
A part of it is due to the way we measure the position of the droplet and some other part is due to numerical errors.
At time 75~ns, we have a sudden small change of $\Vel_y$ and we can only attribute it to the same reasons.

In the next simulations, we use as initial condition the droplet exactly as shown in Fig.~\ref{fig:staticDrop}.
We choose $\alpha=0.1$ and $\STnonad=0.2$.
The larger damping is chosen in order to avoid transients and obtain the essential dynamics in a shorter simulation time.
Furthermore, we now choose $P=0.538$ and this gives $\STFlow=3.72\,{\rm m/sec}$.

We apply a current in the $x$ direction, or $(\STFlow_1,\STFlow_2)=(\STFlow,0)$.
Fig.~\ref{fig:STsnapshots_pi}a shows snapshots of the droplet during the simulation.
The motion is initially complicated with the droplet making a full $\pi$ turn.
For times greater than $150\,{\rm ns}$ a steady-state motion is reached and the velocity has a constant value $7.41\,{\rm m/sec}$ along the direction of the current, in very good agreement with the theoretical prediction $\Vstt=2\STFlow$ given in Eq.~\eqref{eq:vel_rigidMotion}.
We also observe a small component of the velocity perpendicular to the current direction, and we attribute it to numerical errors.
We finally mention that for large values of $\STnonad$ (e.g., $\STnonad = 5\alpha$) the droplet is destroyed while it is moving, by expanding in size.
We continue to present the full set of our simulations before we proceed to give an explanation for the steady state achieved by the droplet.


In the next simulation, we choose a spin current along the $y$ direction, or $(\STFlow_1,\STFlow_2)=(0,\STFlow)$.
Fig.~\ref{fig:STsnapshots_pi}b shows a series of snapshots of the droplet during the motion.
The droplet is initially making a $\pi/2$ turn.
A propagating steady-state is eventually reached with velocity $7.41\,{\rm m/sec}$ in the $y$ direction in very good agreement with the theoretical prediction $\Vstt=2\STFlow$ given in Eq.~\eqref{eq:vel_rigidMotion}.

The important feature shared by all the simulations that we have seen in this section is the common orientation of the skyrmion-antiskyrmion pairs with respect to the direction of motion in the steady state.
Specifically, in both entries of Fig.~\ref{fig:STsnapshots_pi}, the skyrmion-antiskyrmion pair goes in steady state motion only after it rotates in order to achieve the particular orientation.
This is because the steady state achieved is a solitary wave, that is, the propagating droplet is a rigidly propagating solution of the Landau-Lifshitz equation.
Such solutions have well-defined features.
For example, the shape of the solitary wave defines its velocity.
In our case, the particular orientation of the skyrmion-antiskyrmion pair gives a solitary wave velocity in the positive $x$ axis for the case of Figs.~\ref{fig:STsnapshots}, \ref{fig:STsnapshots_pi}a and in the positive $y$ axis in the case of Fig.~\ref{fig:STsnapshots_pi}b.
Exchanging the positions of the skyrmion and the antiskyrmion would invert the direction (sign) of the velocity.
One could conclude that the spin current sets in motion the skyrmion-antiskyrmion droplet revealing its solitary wave character.

The solitary wave character of topologically trivial skyrmionic textures has been studied for the case of a skyrmionium, a topologically trivial, $\Skyrmion=0$, configuration in DM ferromagnets \cite{BogdanovHubert_JMMM1999}.
A static skyrmionium is axially-symmetric and a propagating one is elongated.
A slowly moving skyrmionium presents Newtonian dynamics and a fast moving one (velocity close to the maximum) presents relativistic dynamics Ref.~\cite{KomineasPapanicolaou_PRB2015a}.

\section{A droplet in a disc element}
\label{sec:dropInDot}

We have found static chiral droplets also in the confined geometry of a magnetic disc-shaped element (a magnetic dot).
We apply an energy relaxation algorithm as in Sec.~\ref{sec:dropInFilm}.
This converges and gives a static skyrmion-antiskyrmion droplet for a wide range of parameter values.

\begin{figure}[t]
    \centering
    \includegraphics[width=0.9\columnwidth]{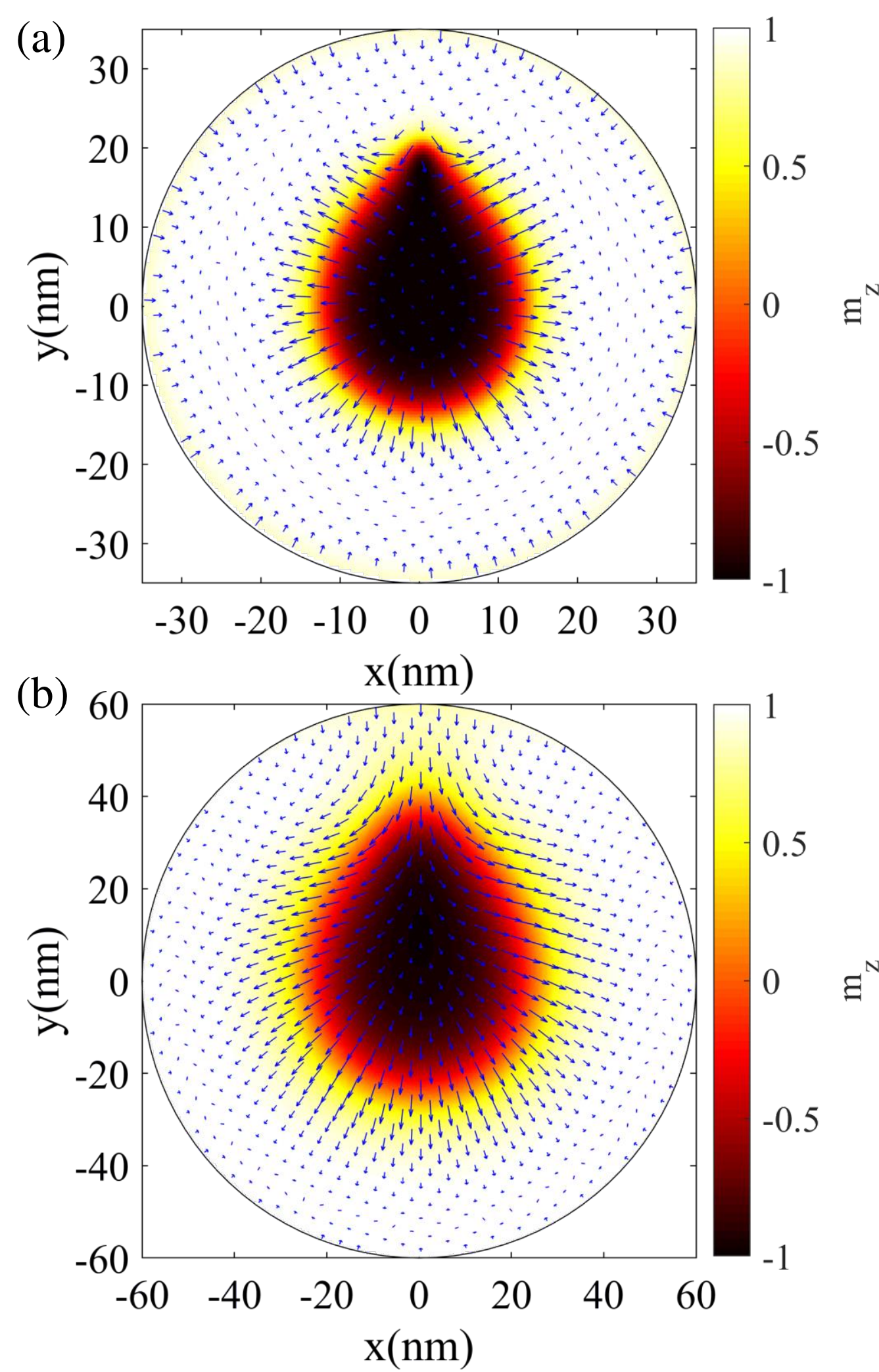}
    \caption{Static skyrmion-antiskyrmion droplets in disc elements with thickness $\thickness=5\,{\rm nm}$.
    (a) The disc diameter is $d=70\,{\rm nm}$.
    The parameter values are as in Table~\ref{tab:parameters}.
    (b) The disc diameter is $d=120\,{\rm nm}$.
    The parameter values are $\DM=1\,{\rm mJ/m^2},\, \Anisotropy=4.43505\times10^5\,{\rm J/m^3}$ (other parameters are as in Table~\ref{tab:parameters}).
    In this case, we have a more extended antiskyrmion part.
    For comparison, we note that a usual skyrmion (with $\Skyrmion=1$) has an approximate radius $15\,{\rm nm}$ for the dot in (a) and $30.5\,{\rm nm}$ for the dot in (b).
    They are thus somewhat larger than the corresponding $\Skyrmion=0$ configurations shown in this figure.}
    \label{fig:dropInDot}
\end{figure}



We find droplets for a thickness $\thickness=0.5\,{\rm nm}$ and for similar parameter values as in the case of a film.
In addition, we were also able to find stable droplet for larger thicknesses.
Fig.~\ref{fig:dropInDot} shows droplets in dots of two different sizes with $\thickness=5\,{\rm nm}$ and for different sets of parameter values.
In Fig.~\ref{fig:dropInDot}a, we have a smaller dot and the parameter values are the same as those used in Fig.~\ref{fig:staticDrop} for the infinite film.
The droplet is stable for a range of parameter values.
For fixed $\dmscaled=0.6086$, the droplet is stable for $3.0\leq D \leq7.6~{\rm mJ/m^2}$.
In Fig.~\ref{fig:dropInDot}b, we have a larger dot and the parameter values (given in the figure caption) correspond to $\epsilon=3.16$ and $\anisotropy=0.00515$.
The value of $\epsilon$ is outside the range for the existence of a skyrmion in a film (when we neglect the long-range part of the magnetostatic field).

In the case of Fig.~\ref{fig:dropInDot}b, the antiskyrmion part is smoother than in all other cases presented in this paper.
This can be attributed to the magnetostatic field originating in the bulk in combination with the DM interaction.
It thus appears that the details of the configuration can be tuned, at least in the case of a droplet in a confined geometry.

The effect of the magnetostatic field due to the confined geometry of a dot is substantial.
The magnetostatic field from the lateral boundaries contributes to stabilizing the configuration.
This phenomenon has already been noted in connection with magnetic bubbles in dots \cite{DruyvesteynSzymczak_PSSA1972,IgnatchenkoMironov_JMMM1993, MoutafisKomineas_PRB2006}.
The effect is verified in the present calculation.

\section{Concluding remarks}
\label{sec:conclusions}

We have found numerically skyrmionic textures in the form of skyrmion-antiskyrmion pairs (droplets) with a skyrmion number $\Skyrmion=0$ in ferromagnets with perpendicular anisotropy and DM interaction.
They exist, in thin films, for a narrow range of parameter values.
The magnetostatic field is crucial for their stability.
Under spin-polarized current, they move along the current exhibiting no Magnus force effect and, thus, their dynamics is different than the dynamics of $\Skyrmion=1$ skyrmions.

The stability of droplets is a numerical finding and we stress that it was not possible to provide a proof for their existence within the Landau-Lifshitz equation including the magnetostatic interaction.
Their robustness is though seen in their behavior under spin currents, where they persist for long times in order to fully reveal their dynamics.

In view of the predicted narrow range of parameters for their stability in infinite films, it would appear as a challenge to observe them experimentally.
Nevertheless, a skyrmion-antiskyrmion has already been observed \cite{JagannathGobelParkin_NatComm2020}.
Skyrmion-antiskyrmion pairs could also be very common as transient (short-lived) states.
We expect that the results of the present paper would help in understanding also such states.

One could consider materials that support antiskyrmions, such as those reported in Ref.~\cite{NayakKumarParkin_Nature2017}.
We have found numerically skyrmion-antiskyrmion droplets also in such systems.
They are very similar to the droplets presented in this paper, except that the skyrmion part is replaced by an antiskyrmion part and vice-versa.

In materials with some special form of DM interaction, such as those studied in Ref.~\cite{HoffmannMelcherBluegel_NatComm2017} (especially in Supplementary Note 1), the skyrmion and the antiskyrmion are both favored.
In such models, a skyrmion-antiskyrmion droplet may have a greater significance.

We have found that skyrmion-antiskyrmion droplets exist for a wider range of parameter values also in confined geometries.
Given the robustness of the $\Skyrmion=0$ droplets in magnetic dots, the present work indicates that some experimental observations of individual skyrmions in magnetic elements might have to be re-examined in order to distinguish whether the observed skyrmions are the symmetric $\Skyrmion=1$ configurations or some sort of $\Skyrmion=0$ textures.
Particular attention should be given to the dynamics of a $\Skyrmion=0$ texture in a dot as this is expected to be different than the rotational dynamics of $\Skyrmion=1$ skyrmions \cite{SisodiaKomineasMuduli_PRB2019}.

\bigskip


\end{document}